\documentclass[preprint]{aastex}

\input{psfig.sty}
\def\etal   {{\it et~al.\/}}

\def\HII    {H~{\rm {II}}}

\begin{document}

\title{Measuring Global Galaxy Metallicities Using Emission Line 
Equivalent Widths}

\author{Henry A. Kobulnicky}
\affil{Department of Physics \& Astronomy \\
University of Wyoming \\ Laramie, WY 82071 
\\ Electronic Mail: chipk@uwyo.edu}

\author{ Andrew C. Phillips, }
\affil{University of California, Santa Cruz \\ Lick Observatory/Board
of Studies in Astronomy \\ Santa Cruz, CA, 95064  \\ 
phillips@ucolick.org}

\author{Revised draft of 28 April 2003 }


\vskip 1.cm

\begin{abstract}

We develop a prescription for estimating the interstellar
medium oxygen abundances of distant star-forming galaxies using the
ratio $EWR_{23}$ formed from the equivalent widths of the [O~II]
$\lambda3727$, [O~III] $\lambda\lambda4959,5007$ and H$\beta$ nebular
emission lines.  This $EWR_{23}$ approach essentially identical to the
widely-used $R_{23}$ method of Pagel \etal\ (1979).  Using data from
three spectroscopic surveys of nearby galaxies, we conclude that the
emission line equivalent width ratios are a good substitute for
emission line flux ratios in galaxies with active star formation.  The
RMS dispersion between $EWR_{23}$ and the reddening-corrected $R_{23}$
values is $\sigma(\log~R_{23})\leq0.08$ dex.  This dispersion is
comparable to the emission-line measurement uncertainties for distant
galaxies in many spectroscopic galaxy surveys, and is smaller than the
uncertainties of $\sigma(O/H)\sim0.15$ dex inherent in strong-line
metallicity calibrations.  Because equivalent width ratios are, to
first order, insentitive to interstellar reddening, emission line
equivalent width ratios may actually be superior to flux ratios when
reddening corrections are not available. The $EWR_{23}$ method
presented here is likely to be most useful for statistically
estimating the mean metallicities for large samples of galaxies to
trace their chemical properties as a function of redshift or
environment.  The approach developed here is used in a companion paper
(Kobulnicky \etal\ 2003) to measure the metallicities of star-forming
galaxies at $z=0.2 - 0.8$ in the Deep Extragalactic Evolutionary Probe
spectroscopic survey of the Groth Strip.

\end{abstract}

\keywords{ISM: abundances --- ISM: \HII\ regions --- 
galaxies: abundances --- 
galaxies: fundamental parameters --- 
galaxies: evolution ---
galaxies: starburst }

\section{Introduction}

Abundances of chemical elements in galaxies are
commonly measured using the emission lines emitted by astrophysical
nebulae (e.g., see Osterbrock 1989).  Recombination lines of hydrogen
(e.g., the Balmer series) or helium and forbidden lines of singly and
doubly ionized carbon, nitrogen, oxygen, neon, and sulfur are among
the strongest observable lines in the visible and ultraviolet
spectrum.  The relative strengths of each emission line, combined with
some knowledge of the temperature and density of the nebulae, provide
information about the relative and absolute abundances of each ion.
Chemical analyses of individual nebulae have been used to test
nucleosynthesis models of stars, the chemical evolutionary history of
galaxies, and nucleosynthesis in the big bang (review by Aller 1990;
Pagel 1998).  Even integrated spectra of entire galaxies may be used
to estimate the overall degree of chemical enrichment of a
galaxy (Kobulnicky, Kennicutt, \& Pizagno 1998).

Observational techniques sometimes permit the relative fluxes
of emission lines from galaxies to be measured with a high degree of
precision.  However, in the current generation of wide-field galaxy
surveys on multi-object spectrographs, flux calibration is frequently
problematic due to unfavorable observing conditions or instrumental
effects such as a variation in system response over the field of view.
For such surveys, it is still desirable to extract chemical
information from the data, where possible.  In this paper we explore
the possibility of estimating mean gas-phase oxygen abundances for
galaxies based the {\it equivalent widths} (rather than fluxes) of
strong Balmer $H\beta$ recombination line and forbidden [O~II] and [O~III]
emission lines.  Our approach is to develop the method from basic
principles, and to test and empirically calibrate it using three
spectroscopic galaxy datasets, and in the end compare the oxygen abundances
derived from traditional line fluxes and flux ratios to results using
only the equivalent widths and equivalent width ratios.  We also
assess the applicability of this method to measuring abundances for a
particular set of intermediate-redshift galaxies observed as part of
the Deep Extragalactic Evolutionary Probe (DEEP) spectroscopic
survey in the Groth Strip
(DGSS) presented in a companion paper, Kobulnicky \etal\ (2003;
Ke03).

\section{Gas-phase Oxygen Abundance Measurements from Emission Lines} 

Osterbrock (1989) reviews the standard techniques for measuring
chemical abundances in astrophysical nebulae from the fluxes and flux
ratios of nebular emission lines.  The most direct and reliable
techniques involve measuring a suite of temperature-sensitive and
density-sensitive line ratios to determine the physical conditions of
the plasma.  In the absence of high signal-to-noise data measuring
temperature--sensitive line ratios (e.g.,
[O~III]~$\lambda$4363/[O~III]~$\lambda$5007), the default diagnostic
for measuring the oxygen abundance of ionized nebulae has become the
ratio of strong oxygen emission lines $R_{23}\equiv
(I_{3727}+I_{4959}+I_{5007})/I_{H\beta}$ (Pagel \etal\ 1979).
Subsequently, many authors have developed formulations relating this
strong line ratio to the gas-phase oxygen abundance (e.g., Edmunds \&
Pagel 1984; McCall, Rybski \& Shields 1985; Dopita \& Evans 1986;
McGaugh 1991).  Other authors (Kobulnicky, Kennicutt, \& Pizagno 1998;
KKP) have shown that this approach is valid even when applied to
galaxies as a whole.  Even when the signal to noise ratio of these
lines is as low as 8:1 or when a spectroscopic observation encompasses
a range of metallicities and ionization conditions within a galaxy,
this ratio provides a rough, but reliable, estimate of the mean
gas-phase oxygen abundance when used in conjunction with the
appropriate calibration relating $R_{23}$ to $O/H$.  At this signal-to
noise ratio, the associated uncertainty on the oxygen abundance due to
line measurement error is typically comparable to the uncertainty due
to the calibration of the $R_{23}$ vs. O/H relationship: $\sim0.15$
dex.  In this paper we show that the uncertainty introduced by
substituting emission line equivalent widths is less than these other
sources of uncertainty, and thereby establish a new method for
measuring interstellar medium oxygen abundances.

\section{Justification of Equivalent Width Method}

The original $R_{23}$ method involves line ratios, so we consider the
effects of using equivalent width ratios in place of flux ratios.  For
generality, we include the effects of reddening in the discussion.  First,
the equivalent width can be written as
\begin{equation}
 W_{\lambda} = \frac{F_{\lambda 1}} {F_{C\lambda 1}}, 
\end{equation}
where $F_{\lambda}$ represents the line flux and $F_{C\lambda}$ is the
underlying continuum flux.  The fluxes do not need to be calibrated, since
the calibration affects both quantities equally.
The dereddened, calibrated flux value for a line, $I_\lambda$,
or continuum, $F_\lambda^0$, is given by
\begin{equation}
 I_{\lambda} = F^0_{\lambda} = F_{\lambda} \times 10^{c(1+f(\lambda))},
\end{equation}
where $c$ is the logarithmic attenuation at $H\beta$ and $f(\lambda)$
is a function describing the reddening curve, according to the prescription
of Seaton (1979).  Combining these leads to a general line ratio formulation of
\begin{equation}
 \frac {I_{\lambda 1}} {I_{\lambda 2}} = 
   \frac {W_{\lambda 1}} {W_{\lambda 2}} \frac {F_{C\lambda 1}} {F_{C\lambda 2}}
    10^{c(f(\lambda 1) - f(\lambda 2))} =
    \frac {W_{\lambda 1}} {W_{\lambda 2}} \alpha,
\end{equation}

where we group all the continuum and reddening terms into the factor
$\alpha$.
Note that the continuum will be attenuated by an amount characterized by
$c^*$, and reddened by (presumably) the same reddening law.  Thus,
\begin{equation}
 \alpha = \frac {F^0_{C\lambda 1}} {F^0_{C\lambda 2}} 
                 10^{(c - c^*)(f(\lambda 1) - f(\lambda 2))}.     
\end{equation}

The factor $\alpha$ contains two unknowns, the ratio of the
dereddened continuum fluxes, which depends on the underlying stellar
population, and the difference in the attentuation of the emitting gas
and the continuum light.  Note that values of $c$ can range from 0 to
quite large for individual HII regions, although large values are
severely deweighted in the average over an entire galaxy.  Typical
derived values are in the range 0--1, with a median value around 0.3
(McCall \etal\ 1985; Oloffson 1995).  Values of $c^*$ can be estimated
from the relation $c^* = 1.47 E_{B-V}$ from Seaton.  For $E_{B-V}$ in a
typical range of 0--0.2, $c^* \sim$ 0--0.3. We would expect some
correlation of $c$ and $c^*$, and also we expect ($c-c^*$) $\geq$ 0.

Let us now consider the specific line ratios of interest for the $R_{23}$
method.  The first is $I_{[O~III]} / I_{H\beta}$.  For the purposes
of reddening, we adopt the wavelength of the stronger [O~III] $\lambda$5007 
line.  The value of $\alpha$ in this case is
\begin{equation}
\alpha = \frac {F^0_{C5007}} {F^0_{C4861}} {10^{(c - c^*)(-0.034)}}.
\end{equation}

Due to the proximity of these emission lines in wavelength, we expect
the flux ratios of the underlying stellar population to be very close to
unity. Similarly, for realistic $0 < c-c^* < 1$, the reddening factor
will be 1 to 0.92.  Thus, for this line ratio,
$\alpha \simeq 1$, and we will ignore it in further discussion.  
To a very good approximation,
\begin{equation}
 \frac {I_{O[III]}} {I_{H\beta}} = \frac {W_{O[III]}} {W_{H\beta}} .
\end{equation}

The second ratio, $I_{O[II]} / I_{H\beta}$, is more problematic; for these
lines,
\begin{equation}
 \alpha = \frac {F^0_{C3727}} {F^0_{C4861}} 10^{(c - c^*)(0.255)}.
\end{equation}
In principle, we could estimate $\alpha$ from a detailed spectral
analysis of the flux-calibrated, integrated stellar spectrum and
Balmer line ratios, but for faint, galaxies with redshifts $z > 0.3$
we will have neither the signal-to-noise nor access to the H$\alpha$
line in order to do this.  In practice, we need to adopt a constant
average value of $\alpha$.

Galaxy light tends to be dominated by A main sequence and G and K
giants (e.g., Morgan \& Mayall 1957;
Pritchet 1977; Kobulnicky \& Gebhardt 2000), and assuming an underlying
stellar spectrum composed of a linear combination of these two
spectral types implies that the dereddened flux ratio of the continuum
ranges from $\sim$1 for late-B stars to $\sim$0.4 for mid-G giants to
$\sim$0.2 for early-K giants (although galaxies whose $\lambda$3727
flux is dominated by K stars is unlikely to have any line emission
from star formation).  Examining the spectra in Kennicutt (1992b)
shows the $\lambda$3727-to-$\lambda$4861 ratio of the continuum fluxes
in galaxies with obvious emission lines ranging from $\sim$0.4--1.0;
if these were dereddened the range would shift upwards and might
possibly narrow somewhat.

On the other hand, the reddening correction ranges from 1 to 1.8 for
realistic values of $c$ and $c^*$, with a typical likely difference $c
- c^* \sim 0.3$ giving a reddening factor of $\sim$1.2.  Combining the
two factors leads to an expected value of $\alpha \sim 0.84 \pm
0.3$.  This average value for $\alpha$ differs from unity by less
than 0.1 dex.

We find the $R_{23}$ measure can now be expressed
\begin{equation}
R_{23} = log \frac {I_{[OII]} + I_{[OIII]}} {I_{H\beta}}
	  = log \frac {\alpha W_{[OII]} + W_{[OIII]}} {W_{H\beta}}.
\end{equation}
Not surprisingly, the $R_{23}$ measured using equivalent widths and a
constant value of $\alpha$ will be most in error when [OII] dominates O[III].

The final line ratio of interest is the ionization parameter, and it is easy
to see that
\begin{equation}
\frac{I_{[OIII]}} {I_{[OII]}} \simeq \frac{W_{[OIII]}} {\alpha W_{[OII]}}.
\end{equation}
This ratio is probably the easiest means to empirically estimate an
average value for $\alpha$.


\section{Data Selection and Analysis}

The thesis of this paper is that even when the line {\it flux} ratio
$R_{23}$ is not
available, a corresponding ratio of {\it equivalent widths} can
still provide an estimate of the oxygen abundance.  This ratio of
equivalent widths is given by

\begin{equation}
EWR_{23} \equiv {{EW([O~II]\lambda3727) + EW([O~III]\lambda4959) +
	EW([O~III]\lambda5007) }\over{EW(H\beta)} }.
\end{equation}

\noindent At the simplest level, $EWR_{23}$ contains neither
corrections for interstellar reddening nor absorption by atmospheres of
the stellar population. To test the suitability of $EWR_{23}$ in place
of $R_{23}$ we compiled three sets of spatially-integrated (i.e.,
global) emission-line spectra of nearby galaxies.  The first set
consists of 16 objects from the 55-object spectroscopic galaxy atlas
of Kennicutt (1992a,b; K92) plus six additional emission-line objects from
KKP.  We refer to this sample as the K92+ sample.  The K92+
spectra are produced by drifting a longslit across each galaxy and have
spectral resolutions of 5-7 \AA\ (K92) and 3 \AA\ (KKP).
The 16 K92 galaxies in our subsample are the
strongest emission-line objects where global metallicity measurements
are possible.  The full set of K92 galaxies represents a range of
morphological types from Sa to Im, but it includes only the bright
galaxies of each type.  As a second local sample, we culled 98 
objects with measurable $H\beta$, [O~III] and [O~II] emission
lines from the 198-galaxy Nearby Field Galaxy Survey (Jansen \etal\
2000a,b; NFGS).  The full NFGS sample is selected from the CfA
redshift catalog (Huchra \etal\ 1983).  These spectra have a
resolution of 6 \AA\ and include a larger range of luminosity
($-14<M_B<-22$) and surface brightnesses than K92+ while spanning
morphological types.  The K92+ objects have a higher fraction of
star-forming galaxies (objects with strong emission lines) compared to
the NFGS sample.  As a third local sample, we used emission-line
selected galaxies in the Kitt Peak National Observatory Spectroscopic
Survey (KISS; Salzer
\etal\ 2000).  KISS is a large-area objective prism survey of local
($z<0.09$) galaxies selected by strong $H\alpha$ emission, and thus,
preferentially contains objects with active star formation.  From a
list of $\sim$519 KISS galaxies with high-quality slit spectroscopy
(Melbourne \& Salzer 2002), we removed 396 galaxies lacking [O~II]
measurements, leaving 123 galaxies.  This analysis makes use of KISS
line strengths obtained during followup longslit spectroscopy with
5--8 \AA\ resolution (Salzer \etal\ 2003).

In summary, the 22 K92+, 98 NFGS, and 123 KISS galaxies were selected
from their larger parent samples because of strong emission lines
suitable for nebular metallicity measurement.  Following KKP, we chose
only galaxies with detectable [O~II]$\lambda$3727,
[O~III]$\lambda$4959, [O~III]$\lambda$5007, and H$\beta$ emission
lines.  Only galaxies where all four emission lines were measured with
a S/N of 8:1 or greater were retained.  This selection criterion
preferentially includes galaxies with high equivalent width lines, but
it also includes galaxies with low equivalent width lines where the
continuum is smooth and well-measured.

For each galaxy in the K92+ sample we measured the
emission-line fluxes and equivalent widths manually using Gaussian
fits.  For the NFGS and KISS surveys, we adopted the published
emission-line fluxes and equivalent widths.  Dereddened emission-line
fluxes were computed for all three samples by comparing the observed
$F_{H\alpha}/F_{H\beta}$ ratios to theoretical ratios.\footnote{
$I_{H\alpha}/I_{H\beta}$ = 2.75-2.86 for wide temperature range,
e.g., (Hummer \& Storey 1987).  Here we assumed fixed electron
temperature of 12,000 K so that $I_{H\alpha}/I_{H\beta}$ = 2.85.}  
The line fluxes 
are dereddened using the law of Seaton (1979) as parameterized
by Howarth (1983) and as described in Kobulnicky \& Skillman (1996).  We
did not correct the Balmer emission lines for underlying stellar
absorption. The effects of Balmer absorption by the stellar 
population are discussed in Section~5.3

While the K92+KKP sample consists entirely of starforming galaxies,
several low-level AGN are known to exist in the NFGS.  These four
objects are not included in the our subsample.  The 123 KISS galaxies
included here do not contain any AGN, as they were selected to be
conventional starforming galaxies based on analysis high-quality
spectroscopic observations (Salzer 2003).  In any case, the presence
of AGN among the samples would not have a significant bearing on the
results of this paper since we are interested in comparing observable
properties of emission lines rather than deriving physical quantities
such as density or metallicity which are sensitive to the nature of
the ionizing source.

\section{Analysis of Emission Line Quantities}

\subsection{[O~II] and [O~III] Equivalent Widths versus Fluxes}

\begin{figure}
\plotone{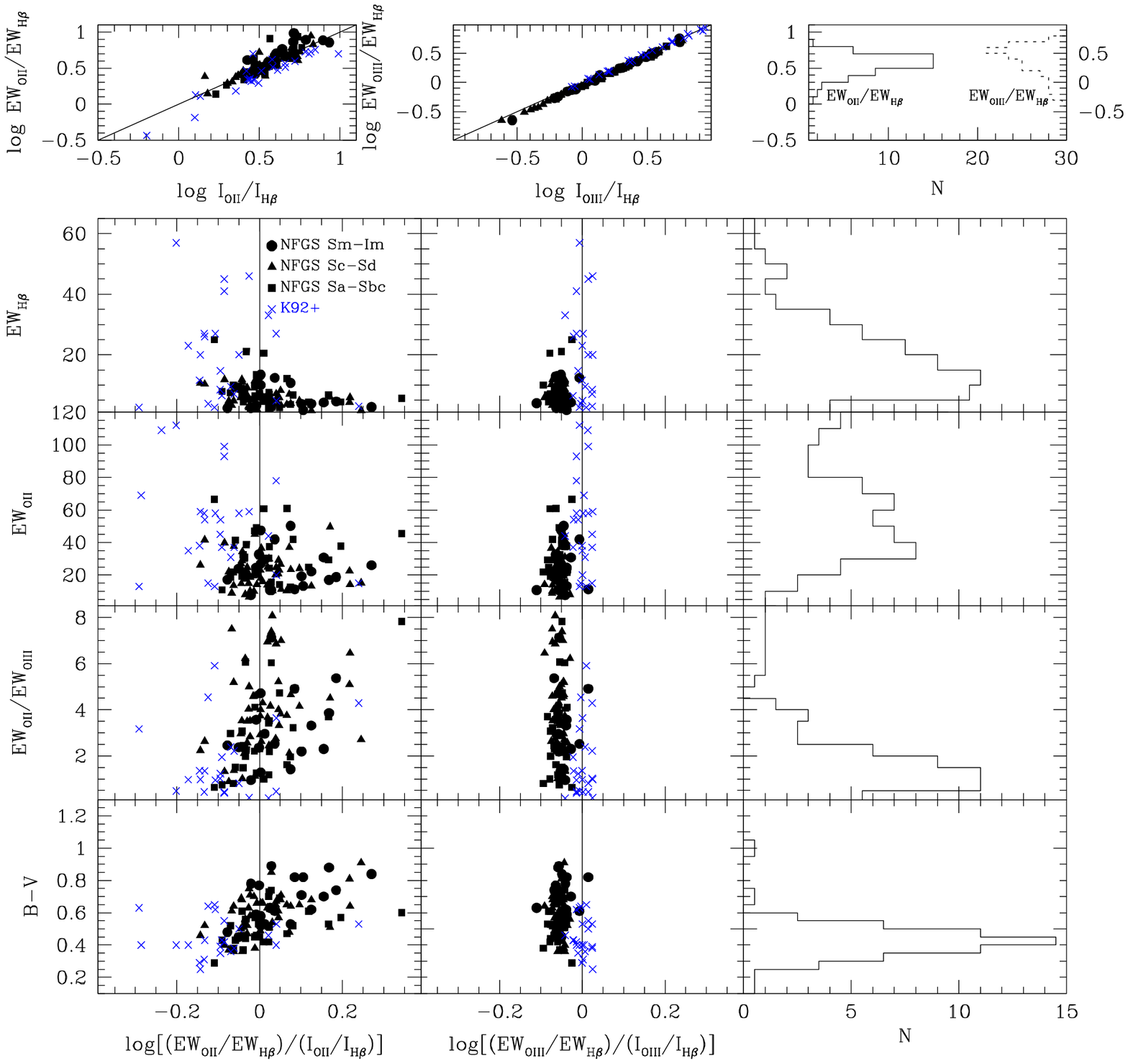}
\figcaption[EWtest3.cps] {Comparison of dereddened emission line fluxes 
and flux ratios
to emission line equivalent widths and EW ratios for galaxies from
Kennicutt (1992a,b) and Jansen \etal\ (2000a,b; NFGS).  Upper panels
show the [O~II]/H$\beta$ flux ratios and EW ratios with a line
illustrating the 1-to-1 correspondence.  Lower panels show residuals
from the 1-to-1 relation as a function of equivalent width and galaxy
color.  There is generally a strong correlation between flux ratios
and EW ratios.  Panels showing systematic residuals are discussed in
the text.  Histograms at right indicate the distribution of emission
line equivalent widths and colors for 66 DGSS galaxies analyzed in
Kobulnicky \etal\ (2003).
\label{EWtest3} }
\end{figure}

Figure~\ref{EWtest3} compares the oxygen and hydrogen emission-line
flux ratios to equivalent width ratios as a function of emission line
strength and $B-V$ color for the K92+ and NFGS samples.  Solid
symbols denote the Nearby Field Galaxy Sample while crosses denote the
K92+ galaxies.  The upper left panel compares the ratio of
dereddened [O~II] to $H\beta$ fluxes, $I_{[O~II]}/I_{H\beta}$, versus
the ratio of [O~II] to $H\beta$ equivalent widths,
$EW_{[O~II]}/EW_{H\beta}$.  A solid line marks the 1-to-1
correspondence.  There is a good correlation between the two
quantities, indicating that strong-line equivalent widths are a good
surrogate for dereddened line fluxes. The RMS deviation from the 1-to-1
correspondence is $\sigma(log[ EW_{[O~II]}/EW_{H\beta}]) = 0.11$ dex
for the combined K92+NFGS samples.  Similarly, the panel at top
center shows the $I_{[O~III]}/I_{H\beta}$ versus
$EW_{[O~III]}/EW_{H\beta}$ ratios and indicates that equivalent widths
are a good surrogate for dereddened line fluxes.  The RMS deviation
from the 1-to-1 correspondence is $\sigma(log[EW_{[O~III]}/EW_{H\beta}])
= 0.05$ dex for the combined K92+NFGS samples.
The panel
at upper right shows the histogram of $EW_{[O~II]}/EW_{H\beta}$ and
$EW_{[O~III]}/EW_{H\beta}$ distributions for the DGSS galaxies (Ke03)
galaxies.  The ratios observed in Groth Strip galaxies fall within the
range where the relations between flux ratios and EW ratios are
well-behaved, suggesting that the $EWR_{23}$ approach could be used to
estimate oxygen abundances of this sample.

The lower panels of Figure~\ref{EWtest3} show residuals from the
1-to-1 line as a function of EW and galaxy color.  The K92+
galaxies have very small residuals in the middle column which compares
[O~III] to $H\beta$ ratios.  The excellent correspondence of [O~III]
fluxes to equivalent widths may be easily understood since the [O~III]
$\lambda\lambda$4959,5007 lines are close in wavelength to H$\beta$ so
that neither changes in the underlying galaxy continuum light nor
relative extinction will alter this ratio. The K92+ and NFGS
galaxies have small, slightly-systematic residuals in the left column
which compares the [O~II] to $H\beta$ ratios.  We attribute the larger
residuals in the [O~II] comparison in the left column to uncertainties
in the reddening correction, and possibly a systematic overestimate of
the reddening correction to the [O~II] $\lambda$3727 flux due to low
spectral resolution in the K92 atlas.  The NFGS, by comparison, shows
systematic trends in the residuals with B-V color, with $EW_{[O~II]}$
and with $EW_{H\beta}$.  The galaxies with the largest residuals are
also the reddest.  They have very low $H\beta$ equivalent widths, and
they have the largest $EW_{[O~II]}/EW_{[O~III]}$ ratios.  Such objects
may be understood to be the galaxies with the lowest rates of star
formation per unit luminosity and the oldest stellar populations.  For
these objects, the line ratios become increasingly uncertain as the
line strengths become weak and the (uncorrected) effect of underlying
Balmer absorption by the stellar population begins to be a dominant
source of scatter.

The right column of histograms in Figure~\ref{EWtest3} shows the
distribution of colors and EW ratios for galaxies in the DGSS selected
for chemical analysis in our companion paper (Kobulnicky
\etal\ 2003).  The selected DGSS galaxies have strong emission
lines and tend to be most similar to the K92+ and KISS galaxies,
having blue colors, small $EW_{[O~II]}/EW_{[O~III]}$ ratios, and large
$EW_{H\beta}$ in comparison the NFGS.  These characteristics suggest
that, like the K92+ and NGFS galaxies with these properties, the
DGSS galaxies will exhibit a close correlation between the emission
line equivalent width and flux ratios, facilitating chemical abundance
analysis.

\begin{figure}
\plotone{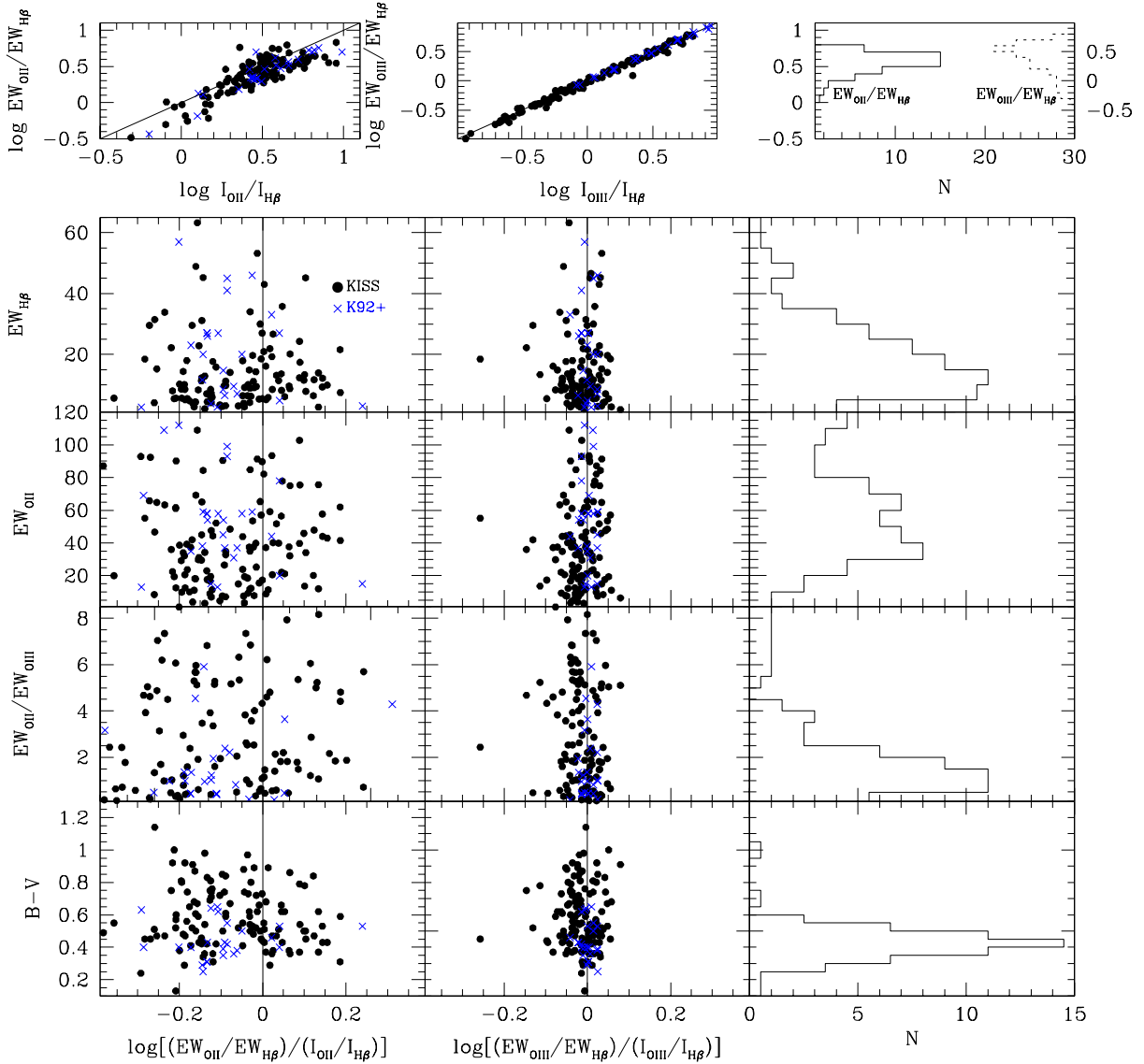}
\figcaption[EWtest3KISS.cps] {Comparison of dereddened emission line fluxes and 
flux ratios to emission line equivalent widths and EW ratios for
galaxies from the K92+ galaxies and the KISS (Salzer \etal\ 2001)
emission line galaxy
survey.  Upper panels show the [O~II]/H$\beta$ flux ratios and EW
ratios with the line illustrating the 1-to-1 correspondence.  Lower
panels show residuals from the 1-to-1 relation as a function of
equivalent width and galaxy color.  There is generally a strong
correlation between flux ratios and EW ratios.  Panels showing
systematic residuals are discussed in the text.  Histograms at right
indicate the distribution of emission line equivalent widths and
colors for DGSS galaxies analyzed in Kobulnicky \etal\ (2003).
\label{EWtest3KISS} }
\end{figure}

In Figure~\ref{EWtest3KISS} shows the same comparison as
Figure~\ref{EWtest3} but for the KISS and K92+ samples.  The upper
left panel compares the ratio of dereddened [O~II] to $H\beta$ fluxes,
$I_{[O~II]}/I_{H\beta}$, versus the ratio of [O~II] to $H\beta$
equivalent widths, $EW_{[O~II]}/EW_{H\beta}$.  A solid line marks the
1-to-1 correspondence.  There is a good correlation between the two
quantities, indicating that strong-line equivalent widths are a good
surrogate for dereddened line fluxes. The RMS deviation from the 1-to-1
correspondence is $\sigma(log EW_{[O~II]}/EW_{H\beta}) = 0.15$ dex for
the combined K92+KISS samples.  Similarly, the panel at top center
shows the $I_{[O~III]}/I_{H\beta}$ versus $EW_{[O~III]}/EW_{H\beta}$
ratios and indicates that equivalent widths are a good surrogate for
dereddened line fluxes.  The RMS deviation from the 1-to-1 correspondence
is $\sigma(log[EW_{[O~II]}/EW_{H\beta})] = 0.04$ dex for the combined
K92+KISS samples.

The lower panels of Figure~\ref{EWtest3KISS} show residuals from the
1-to-1 line as a function of EW and galaxy color. The middle column
shows very small and non-systematic residuals, indicating excellent
correspondence between [O~III] equivalent widths and fluxes for all
KISS galaxy colors and line ratios.  However, the left column shows
significant dispersion of 0.15 dex between
$\log~EW_{[O~II]}/EW_{H\beta}$ and $\log~I_{[O~II]}/I_{H\beta}$.  The
residuals are not correlated with galaxy color or EW, suggesting that
measurement errors and/or uncertainties in the reddening correction
are responsible for the dispersion.
 
The right column of histograms in Figure~\ref{EWtest3KISS} shows the
distribution of colors and EW ratios for galaxies in the DGSS selected
for chemical analysis in our companion paper (Kobulnicky
\etal\ 2003). 

\subsection{EW$R_{23}$ versus $R_{23}$}

\begin{figure}
\plotone{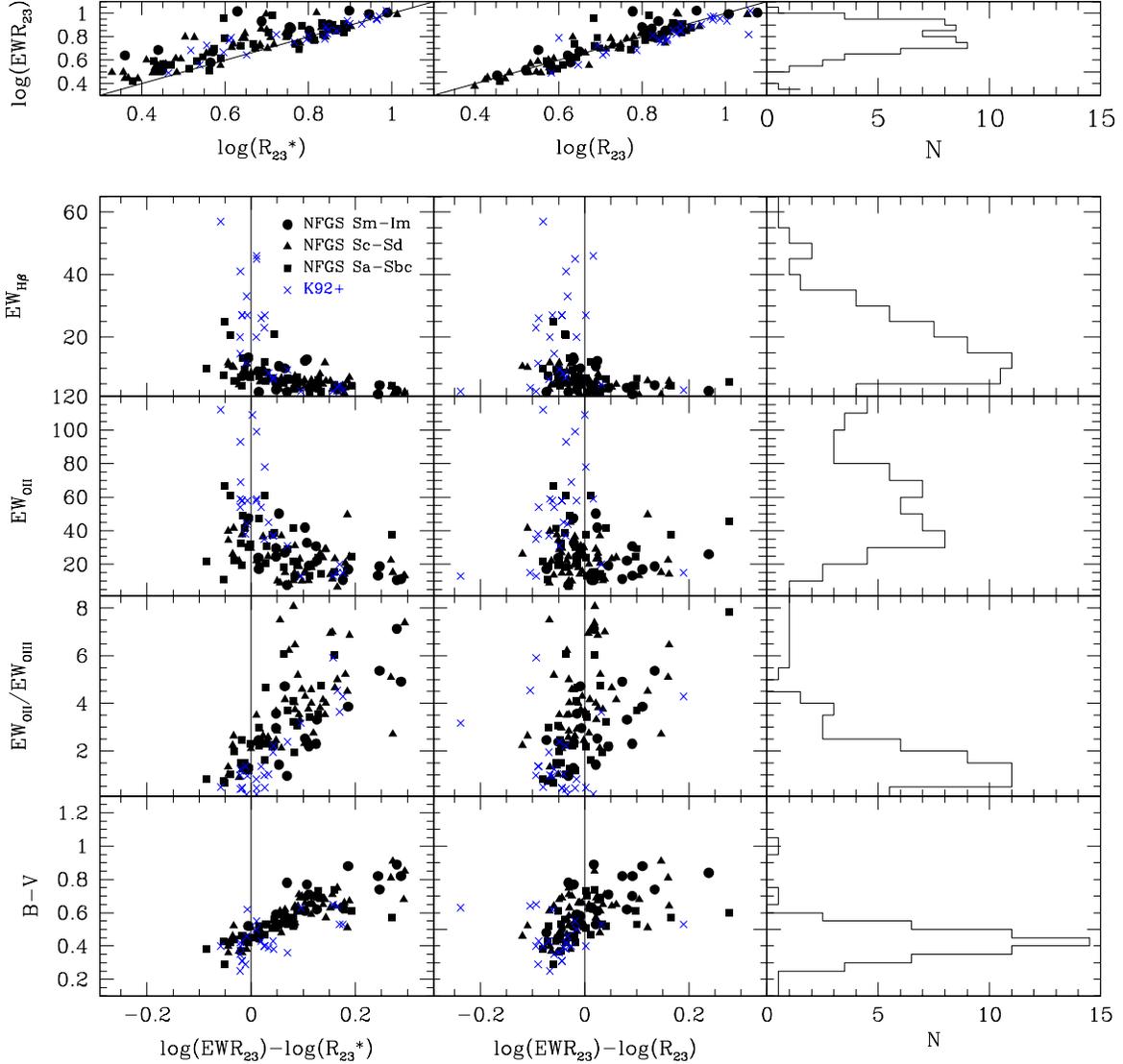}
\figcaption[EWtest2.ps] {
Comparison of the quantity $R_{23}$ and $R_{23}^*$ with $EWR_{23}$.
$R_{23}^*$ is $R_{23}$ without correction for reddening.  The strong
correlation between $R_{23}^*$ and $EWR_{23}$ (upper left panel) is
even stronger for $R_{23}$ and $EWR_{23}$, suggesting that 
oxygen abundances can be estimated from equivalent width ratios at
least as well as from dereddened line fluxes. 
The RMS dispersion from the 1-to-1 relation is
$\sigma(log[R_{23}]) = 0.07$ dex.   Lower panels explore
the residuals in the correlation as a function of line strength, line
ratio, and galaxy color.  {\it left column}: no correction for
extinction or underlying Balmer absorption in either quantity; {\it
middle column}: line fluxes have been corrected for extinction; {\it
right column}: histogram of the distribution of 66 emission-line
galaxies selected for analysis from the DGSS showing that the majority
of galaxies lie in regimes where the $EWR_{23}$ vs. $R_{23}$
correlation is strong and well-behaved.
\label{EWtest} }
\end{figure}

A more direct test of the suitability of emission line equivalent
width ratios for abundance analysis can be achieved by comparing the
quantity EW$R_{23}$ with $R_{23}$ for the same three local galaxy
samples.  Figure~\ref{EWtest} (upper left panel) shows the comparison
between $R_{23}^*$ and $EWR_{23}$ for K92+ and NFGS galaxies
constructed from the raw fluxes and equivalent widths.  $R_{23}^*$
denotes the value of $R_{23}$ {\it without} correcting line fluxes for
reddening.  A solid line illustrates a 1-to-1 correspondence.  The lower
rows in the left column of Figure~\ref{EWtest} show residuals from the
1-to-1 correspondence as a function of $EW_{H\beta}$, $EW_{[O~II]}$,
$EW_{[O~II]}/EW_{[O~III]}$ and galaxy $B-V$ color.  The correlation
between $R_{23}^*$ and $EWR_{23}$ is strong but has considerable
scatter.  Formally, the RMS dispersion from the 1-to-1 relation is
$\sigma(log[R_{23}^*]) = 0.12$ dex for the combined K92+NFGS
samples.

The correlation is strongest for objects in the K92+ sample and for
objects with large values of $R_{23}^*$.  Deviations from 1-to-1 are
greatest for objects in the NFGS sample which have low $EW_{H\beta}$,
high $EW_{[O~II]}/EW_{[O~III]}$ ratios, and red colors.  These
systematic residuals may be understood as a consequence of the lack of
corrections for extinction.  [O~II] $\lambda$3727 is significantly
affected by extinction compared to the H$\beta$ and [O~III] lines.
The measured [O~II] flux is a lower limit to the true unextincted
intensity whereas the measured [O~II] equivalent width should be
unaffected by extinction provided that the extinction toward the gas
and stars are similar\footnote{See Calzetti, Kinney, \&
Storchi-Bergmann (1994) for evidence that this assumption is sometimes
invalid.}.  The strong systematic residuals with $B-V$ color seen in
the lower left panel is a most likely a consequence uncorrected
extinction, since redder galaxies are often those with greater
extinction.

In the middle column of Figure~\ref{EWtest} we show a similar
comparison of $EWR_{23}$ with $R_{23}$, where $R_{23}$ has been
corrected for reddening using the theoretical Balmer decrement.  Here
the correlation is much stronger.  The strong correlation between
galaxy color and residuals seen in left column is now mostly gone,
suggesting that the [O~II] line fluxes have been successfully
corrected for reddening.  The RMS dispersion from the 1-to-1 relation is
$\sigma(log[R_{23}]) = 0.07$ dex for the combined K92+NFGS samples.
The ratio of equivalent widths is a good substitute for the
reddening-corrected $R_{23}$ ratio.  Use of equivalent widths may even
be superior to line ratios if the reddening corrections are not known,
as in the case of galaxies for which the H$\alpha$/H$\beta$ ratio is not
available (typically true for redshifts $z > 0.3$).
In the presence of reddening, the EW of the [O~II] line is unchanged
because both the continuum and line flux are suppressed in equal
amounts if the reddening toward stars and gas is similar.  The
residuals in the lower panels of column 2 are mostly symmetric about
zero, with the largest scatter again occurring for objects with very
low values of EW(H$\beta$).  Some of the systematic residuals
are probably also caused by varying continuum shapes, especially
among the NFGS, which affect the equivalent widths of
the [O~II] lines in a systematic manner which is realted to
galaxy color and the average age of the stellar population.
 In any case, the RMS of 0.07 dex in
$R_{23}$ will often be on the same order as, or even less than the
statistical measurement uncertainties on the strong line equivalent
widths in high-redshift spectroscopic surveys, even when the signal-to-noise 
of the emission line equivalent widths is as low as 8:1.

The right column of Figure~\ref{EWtest} shows the distribution of the
66 galaxies from the DGSS selected for chemical analysis in Kobulnicky
\etal\ (2003).
The range of $EWR_{23}$, $EW_{H\beta}$, $EW_{[O~II]}$, and
$EW_{[O~II]}/EW_{[O~III]}$ among the sample galaxies covers the range
over which the local calibration galaxies show a well-behaved
relationship between $EWR_{23}$ and $R_{23}$.  The good correlation
($\sigma_{23}$=0.08 dex) between $R_{23}$ and $EWR_{23}$ in the local
samples provides confidence that measuring rough ISM oxygen abundances
in large surveys of distant galaxies is feasible even when only equivalent
widths are recorded.

\begin{figure}
\plotone{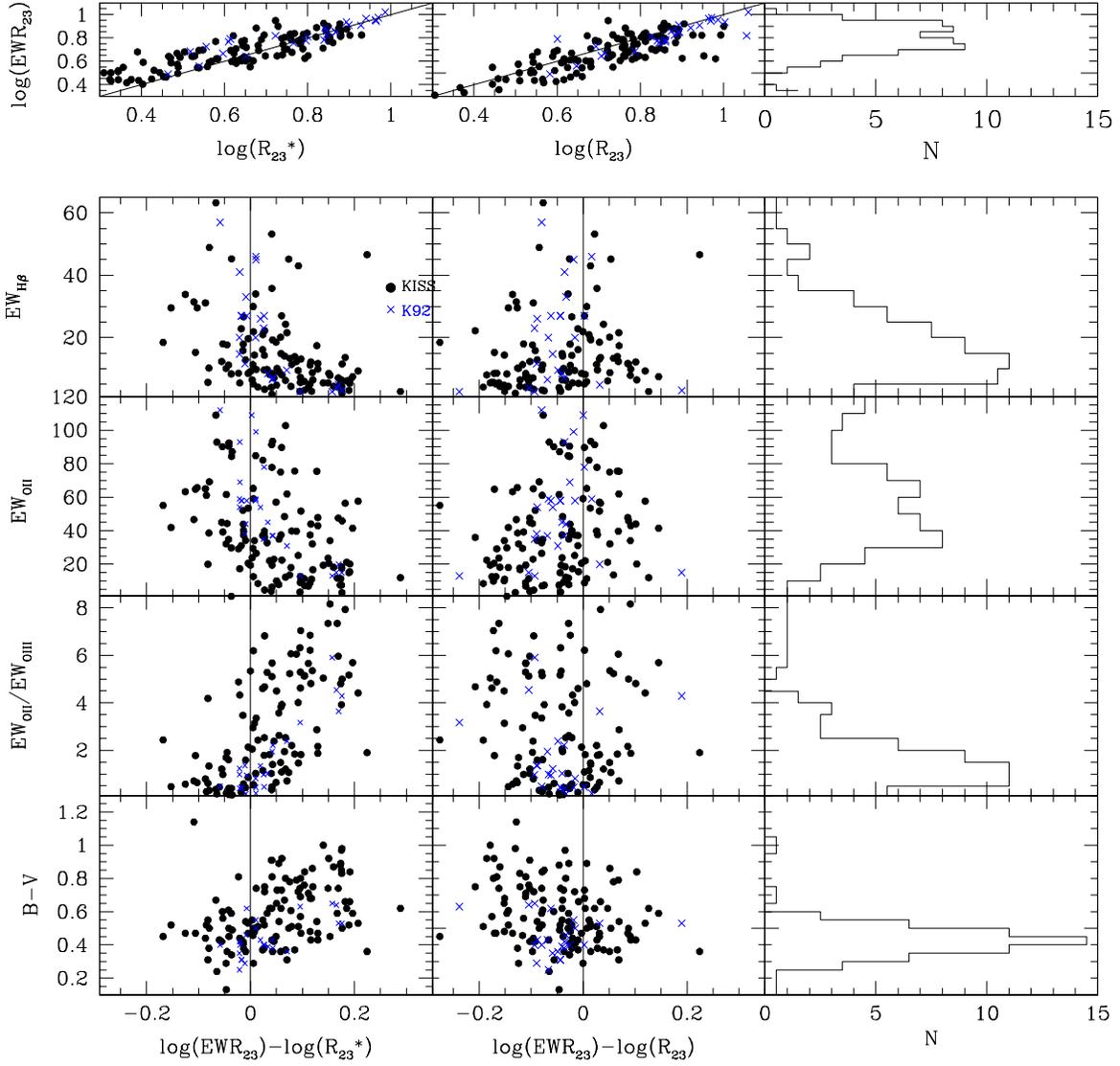}
\figcaption[EWtest2KISS.cps] 
{Comparison of the quantity $R_{23}$ and $R_{23}^*$ with $EWR_{23}$
for the K92+ and KISS galaxy samples. Upper panels show the
excellent correlation between $R_{23}$ and $R_{23}^*$ with $EWR_{23}$.
Lower panels show the residuals from the 1-to-1 correspondence as a
function of line strength and galaxy color.  Residuals are larger, but
less systematic, for the KISS galaxies than for the NFGS and K92+
galaxies.  As in previous figures, histograms in the right column show
the distribution of DGSS galaxies analyzed using this approach in
Ke03.
\label{EWtestKISS} }
\end{figure}

Figure~\ref{EWtestKISS} shows a comparison of EW$R_{23}$ with
$R_{23}^*$ and $R_{23}$ for the K92+ and KISS galaxy samples.  The
left column shows the comparison of EW$R_{23}$ with $R_{23}^*$ and the
associated residuals from the 1-to-1 correspondence.  There is a good
correlation between EW$R_{23}$ with $R_{23}^*$ in the upper left
panel.  The RMS dispersion from the 1-to-1 relation is
$\sigma(log[R_{23}]) = 0.11$ dex for the combined K92+KISS
samples.  The left column shows systematic residuals with galaxy color
and line strength indicating that reddening is significant for some
galaxies.  The center column shows the comparison of EW$R_{23}$ with
$R_{23}$ and the associated residuals as a function of EW and galaxy
color.  The residuals for the KISS galaxies are now slightly smaller
and much less systematic after application of a reddening
correction. The RMS dispersion from the 1-to-1 relation is
$\sigma(log[R_{23}]) = 0.09$ dex for the combined K92+KISS
samples.  The residuals are larger than than for the K92+ and NFGS
objects, but the KISS galaxies do not show the systematics with galaxy
color or EW which the NFGS galaxies exhibit.  The increased scatter in
$R_{23}$ may be traced to the increased scatter in the [O~II] lines
seen in Figure~\ref{EWtest3KISS}.  The lack of systematic residuals is
probably a result of the relative homogeneity of the KISS sample.
KISS galaxies have stronger emission lines and do no include the
diversity of more quiescent galaxies with older stellar populations
found in the NFGS.

\subsubsection{Effects of Stellar Balmer Absorption}

Ideally, the quantity $R_{23}$ from which a metallicity is derived
should be computed using an $H\beta$ line strength which has been
corrected for both interstellar reddening {\it and} absorption by
atmospheres of the underlying stellar population.  In practice, the
amount of underlying absorption is difficult to measure even under
ideal circumstances with high signal-to-noise data.  Spectra of
distant galaxies frequently lack the signal-to-noise necessary to
measure multiple Balmer lines and correct simultaneously for
extinction and Balmer absorption in a self-consistent fashion.  For
galaxies with strong emission lines due to active star formation
(i.e., $EW_{H\beta} >25$ \AA), a correction of a few \AA\ to the
$H\beta$ line will have a small impact on the derived $R_{23}$ or
$EWR_{23}$.  However, in galaxies dominated by older stellar
populations with weak emission lines, $R_{23}$ or $EWR_{23}$ will
depend sensitively on the correction for Balmer absorption.

Until this point in the analysis, we have not made any corrections for
stellar Balmer absorption.  The effect of underlying Balmer absorption
(specifically the amount of absorption in the H$\beta$ line,
$EW_{H\beta}(abs)$ will depress the measured $F_{H\beta}$ and
$EW_{H\beta}$.  This leads to systematically large $R_{23}$ or
$EWR_{23}$, and systematically low oxygen abundances for objects on
the upper (metal-rich) branch of the empirical calibrations.

\begin{figure}
\plotone{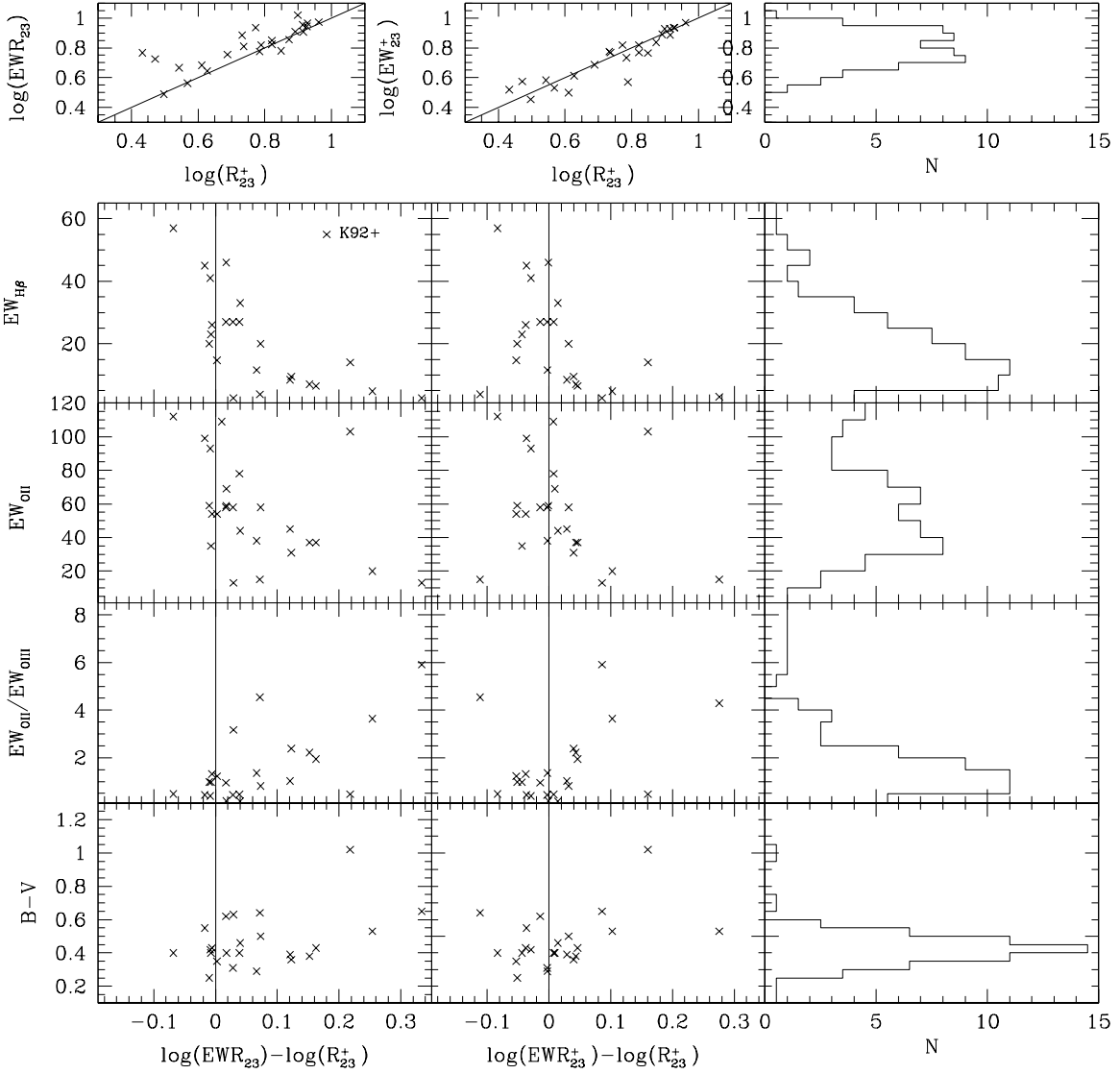}
\figcaption[EWtest4.ps] 
{Comparison of the quantity $R_{23}^+ $ with $EWR_{23}$ and
$EWR_{23}^+$ for the K92+ galaxies only.
$EWR_{23}^+$ is $EWR_{23}$ where $EW_{H\beta}$ has been
increased by 2 \AA\ to account for stellar absorption.
 $R_{23}^+$ is $R_{23}$
with self-consistent corrections for both reddening and stellar Balmer
absorption.  Upper panels show the correlation between $R_{23^+}$ and
with $EWR_{23}$ and $EWR_{23^+}$.  Lower panels show the residuals
from the 1-to-1 correspondence as a function of line strength and galaxy
color.  Galaxies with $ EW_{H\beta}\leq15 $ \AA\ are most seriously
affected by the lack of correction for stellar absorption.
\label{EWtest4} }
\end{figure}

The impact of stellar absorption can be assessed using
Figure~\ref{EWtest4}.  Using the K92+ galaxies, we performed a
self-consistent reddening and stellar absorption correction for each
galaxy.  The upper left panel compares the raw $EWR_{23}$ ratio with
the quantity $R_{23}^+$ which includes the corrections for reddening and stellar
absorption.
The correlation between $EWR_{23}$ and $R_{23}^+$ is
modest, with a dispersion of $\sigma(R_{23}^+)=0.14$ dex and a
systematic offset of 0.06 dex.  The lower panels illustrate the nature
of the residuals as a function of galaxy color and EW.  As might be
expected, galaxies with the lowest $EW_{H\beta}$ are the most deviant,
while galaxies with $EW_{H\beta}>20$ show a much smaller dispersion.
A more logical approach is also to add a correction to $EW_{H\beta}$ for the
underlying stellar absorption, forming a new quantity $EWR_{23}^+$.
Galaxies with the lowest $EW_{H\beta}$ are affected most by this correction.
A 2 \AA\ correction to $EW_{H\beta}$ was chosen because it was
the mean correction needed to produce $R_{23}^+$, and is consistent
with mean corrections found for other galaxies (e.g., McCall \etal\
1985; Olofsson 1995).
The center top panel shows $EWR_{23}^+$ plotted against $R_{23}^+$.
The residuals are now much smaller with
$\sigma(R_{23}^+)=0.08$ dex and a systematic offset of less than 0.01
dex.
In the absence of direct measurements of the
Balmer absorption due to stellar populations, application of a 2 \AA\
blanket correction to $EW_H{\beta}$ appears to be prudent.

\subsection{The Ionization Parameter Quantity [O~III]/[O~II]}

\begin{figure}
\plotone{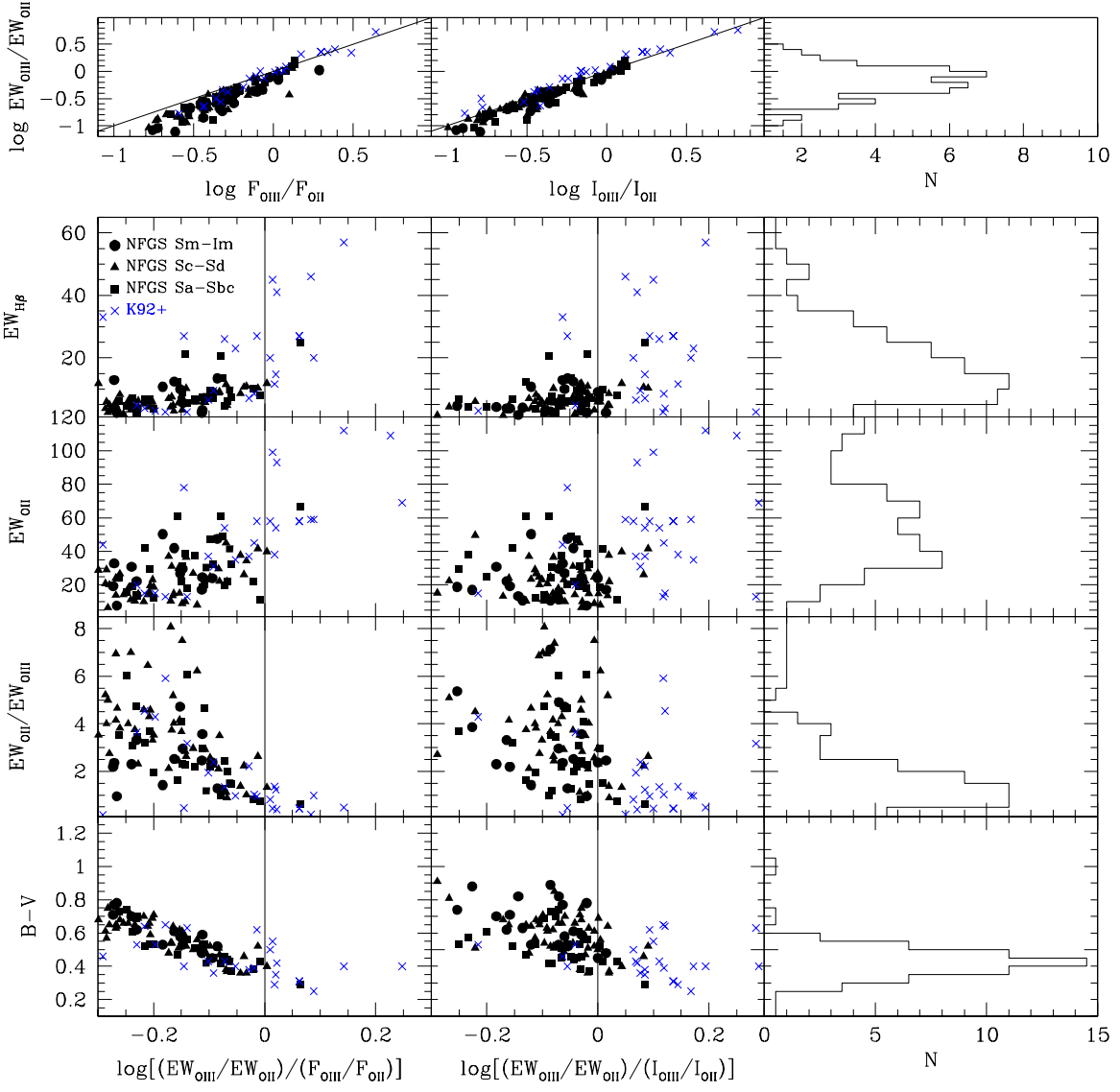}
\figcaption[EWtest5.cps] 
{Comparison of the ionization parameter quantity $log
(EW_{[O~III]}/EW_{[O~II]})$ with $log (F_{[O~III]}/F_{[O~II]})$ and
the dereddened ratio $log (I_{[O~III]}/I_{[O~II]})$ for the K92+
and NFGS galaxies.  Lower panels show the residuals from the 1-to-1
correspondence as a function of line strength and galaxy color.
\label{EWtest5} }
\end{figure}

Modern calibrations relating $R_{23}$ to oxygen abundance include a
measure of the ionization parameter, such as $O_{32}\equiv
(F_{4959}+F_{5007})/F_{3727} = F_{[O~III]}/F_{[O~II]}$ as a second
parameter (e.g., McGaugh 1991; Pilyugin 2001).  We test the
suitability of using a quantity $EW_{[O~III]}/EW_{[O~II]}$ in place of
$F_{[O~III]}/F_{[O~II]}$ in Figure~\ref{EWtest5}.  K92+ galaxies
appear as crosses and the NFGS galaxies appear as solid symbols.  The
upper left panel shows the correlation between
$EW_{[O~III]}/EW_{[O~II]}$ and $F_{[O~III]}/F_{[O~II]}$, where $F$
denotes line strengths that have not been corrected for reddening.
Although the correlation is strong, the residuals in the lower rows
are large and strongly systematic with galaxy color and emission line
$EW$.  The top middle panel shows the correlation between
$EW_{[O~III]}/EW_{[O~II]}$ and $I_{[O~III]}/I_{[O~II]}$, where $I$
denotes line strengths that have been corrected for reddening.  The
correlation is now much stronger and less systematic.  The RMS
dispersion is $\sigma(I_{[O~III]}/I_{[O~II]})=0.12$ dex.  The mean
residuals of the two samples are systematic in different directions.
The residuals of the K92+ galaxies suggest a systematic reddening
overcorrection of the [O~II] $\lambda$3727 line.  The residuals for
the NFGS galaxies are still systematic with galaxy color and suggest a
reddening undercorrection for the reddest galaxies.  The two samples
have been analyzed identically, and we have no explanation for the
apparent differences.  A more rigorous comparison would require
re-measuring all of the line strengths and equivalent widths to ensure
uniform treatment of the samples.  However, electronic spectra for the
NFGS and KISS galaxies are not publicly available.

\begin{figure}
\plotone{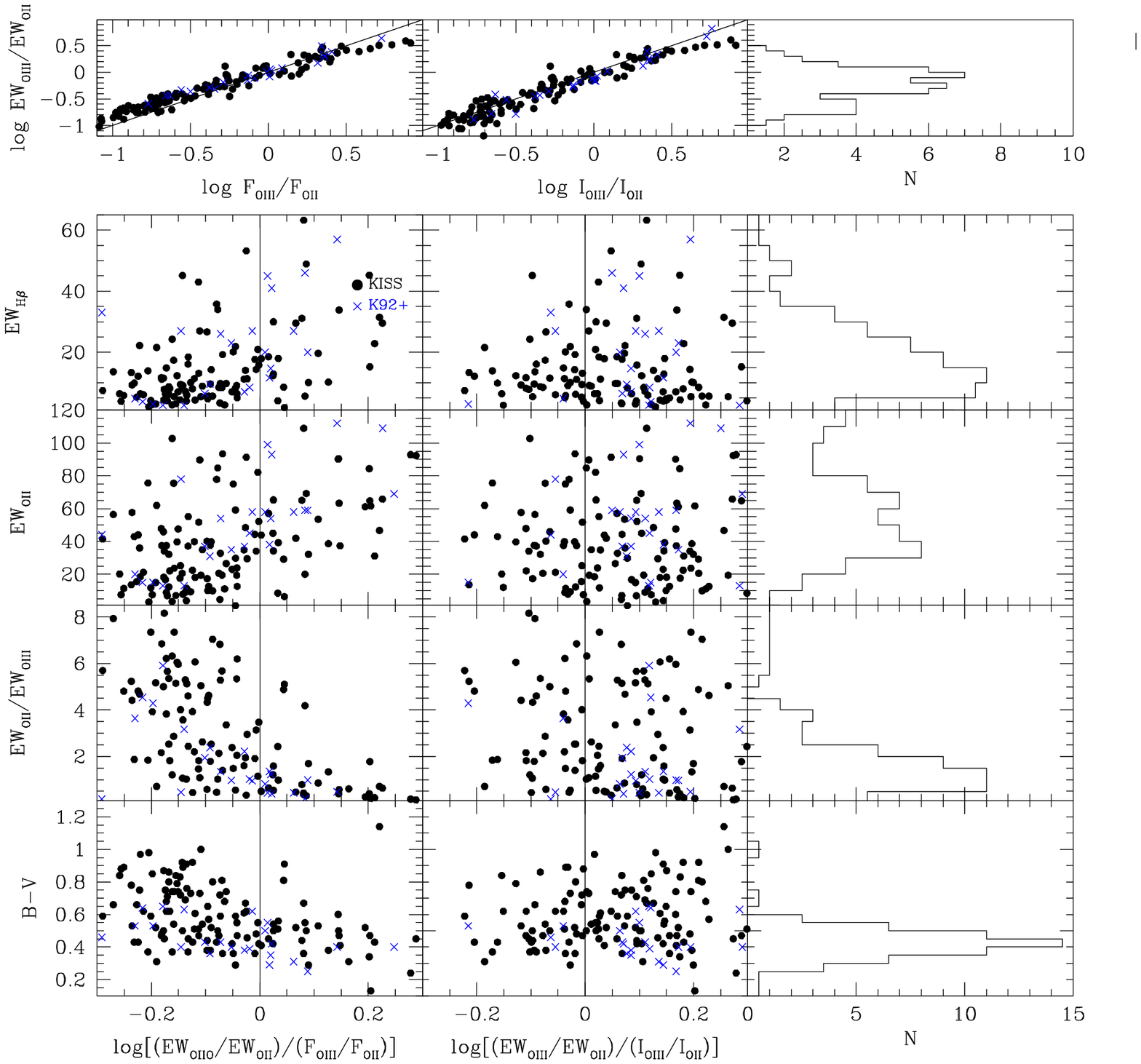}
\figcaption[EWtest5KISS.cps] 
{Comparison of the ionization parameter quantity $log
(EW_{[O~III]}/EW_{[O~II]})$ with $log (F_{[O~III]}/F_{[O~II]})$ and
the dereddened ratio $log (I_{[O~III]}/I_{[O~II]})$ for the K92+
and KISS galaxies.  Lower panels show the residuals from the 1-to-1
correspondence as a function of line strength and galaxy color.
\label{EWtest5KISS} }
\end{figure}

In Figure~\ref{EWtest5KISS} we show the $EW_{[O~III]}/EW_{[O~II]}$
versus $F_{[O~III]}/F_{[O~II]}$ and $I_{[O~III]}/I_{[O~II]}$
comparisons for the KISS galaxies.  The residuals for both the left
and center columns are $\sigma(F_{[O~III]}/F_{[O~II]})=0.15$ dex and
$\sigma(I_{[O~III]}/I_{[O~II]})=0.14$ dex.  The lack of improvement
with reddening correction is undoubtedly traceable to the large,
apparently random, dispersion in the [O~II] line strengths and EWs of
the KISS spectra, as noted in previous figures.  

In summary, the $EW_{[O~III]}/EW_{[O~II]}$ ratios could be used as a
surrogate for $F_{[O~III]}/F_{[O~II]}$ ratios as a indicator of the
ionization parameter.  The residuals are on the order of $0.12-0.15$
dex and are sometimes strongly systematic with galaxy color.  The mean
of the residuals varies considerably from -0.08 dex to 0.12 dex in
$EW_{[O~III]}/EW_{[O~II]}$ depending on the galaxy sample under study.

\section{Discussion and Conclusions}

The ratios of the equivalent widths of strong oxygen and hydrogen
emission lines from the ionized component of distant galaxies can be
used as a measure of the global ISM metallicity via the substitution
of $EWR_{23}$ for $R_{23}$. We recommend the use of $EWR_{23}^+$ where
the $EW_{H\beta}$ has been corrected for stellar Balmer absorption
assuming a mean correction of 2 \AA.  The typical dispersion from the
1-to-1 relation between either $EWR_{23}$ or $EWR_{23}^+$ and the
canonical reddening and absorption-corrected $R_{23}$ ratio is
$\sigma(R_{23})=0.08$ dex.  Residuals are somewhat smaller
($\sigma=0.05$ dex) for galaxies with the largest emission-line
equivalent widths (i.e., those having the largest rates of star
formation per unit luminosity).  The additional uncertainty introduced
by translating a set of measured equivalent widths into the
traditional $R_{23}$ flux ratio diagnostic is comparable to or less
than the typical observational line measurement uncertainties and
systematic errors in the $R_{23}$ to O/H calibration which run 0.15
dex in O/H (e.g., Kobulnicky, Kennicutt \& Pizagno 1999 for a more
detailed discussion of the error budget).  We anticipate that the
method tested here will be useful for performing rough chemical
abundance estimates in large high-redshift galaxy samples. 
The approach described here
will be most useful in a statistical sense when large numbers of
objects are available for study.  Possible applications include
understanding the overall chemical evolution of star forming galaxies
over large intervals of cosmic time (Kobulnicky \etal\ 2003) or
assessing the impact of cluster environment and the intracluster
medium on the chemical properties of the ISM within galaxies (e.g.,
Skillman \etal\ 1996).

\acknowledgments

John Salzer helped make this paper possible by providing the KISS data
in tabulated form and by commenting on the manuscript.  
Detailed comments by an anonymous referee greatly improved the 
manuscript.  We thank
Shiela Kannappan for a helpful discussion about the NFGS and Matt
Bershady for scientific inspiration.  H.~A.~K was supported by NASA
through grant \#HF-01090.01-97A awarded by the Space Telescope Science
Institute which is operated by the Association of Universities for
Research in Astronomy, Inc. for NASA under contract NAS 5-26555.  This
work was also made possible by funding from the National Science
Foundation through grant AST-9529098, and NASA through
NRA-00-01-LTSA-052.

\clearpage

\end{document}